# Generation of GeV positron and γ-photon beams with controllable angular momentum by intense lasers

**Authors:**

Xing-Long Zhu[1, 2], Tong-Pu Yu[3], Min Chen[1, 2], Su-Ming Weng[1, 2] and Zheng-Ming Sheng[1, 2, 4, 5†]

**Affiliations:**

[1] Key Laboratory for Laser Plasmas (Ministry of Education), School of Physics and Astronomy, Shanghai Jiao Tong University, Shanghai 200240, China

[2] Collaborative Innovation Center of IFSA (CICIFSA), Shanghai Jiao Tong University, Shanghai 200240, China

[3] Department of Physics, National University of Defense Technology, Changsha 410073, China

[4] SUPA, Department of Physics, University of Strathclyde, Glasgow G4 0NG, UK

[5] Tsung-Dao Lee Institute, Shanghai 200240, China

[†]e-mail: zmsheng@sjtu.edu.cn

**ABSTRACT**

**Although several laser-plasma-based methods have been proposed for generating energetic electrons, positrons and γ-photons, manipulation of their microstructures is still challenging, and their angular momentum control has not yet been achieved. Here, we present and numerically demonstrate an all-optical scheme to generate bright GeV γ-photon and positron beams with controllable angular momentum by use of two counter-propagating circularly-polarized lasers in a near-critical-density plasma. The plasma acts as a 'switching medium', where the trapped electrons first obtain angular momentum from the drive laser pulse and then transfer it to the γ-photons via nonlinear Compton scattering. Further through the multiphoton Breit-Wheeler process, dense energetic positron beams are efficiently generated, whose angular momentum can be well controlled by laser-plasma interactions. This opens up a promising and feasible way to produce ultra-bright GeV γ-photons and positron beams with desirable angular momentum for a wide range of scientific research and applications.**

# INTRODUCTION

Positrons—as the antiparticles of electrons—are relevant to a wide variety of fundamental and practical applications[1-3], ranging from studies of astrophysics and particle physics to radiography in material science and medical application. However, it is very difficult to produce dense energetic positrons in laboratories[4], even though they may exist naturally in massive astrophysical objects like pulsars and black holes[2] and are likely to form gamma-ray bursts[5]. Recently, significant efforts have been devoted to producing positron beams by use of intense laser pulses. For example, with laser-produced energetic electrons interacting with high-Z metals, positrons are generated via the Bethe-Heitler process[6] with typical density of $\sim 10^{16} \text{cm}^{-3}$ and average energy of several MeV[7-10]. The abundant positrons with a higher energy and extreme density are particularly required for diverse areas of applications[1-5], such as exploring energetic astrophysical events, nonlinear quantum systems, fundamental pair plasma physics, and so on. The upcoming multi-PW lasers[11, 12] will allow one to access the light intensity exceeding $10^{23} \text{W cm}^{-2}$, which pushes light-matter interactions to the exotic QED regime[1, 12-14], including high-energy γ-photon emission and dense positron production. Under such laser conditions, both theoretical[15-17] and numerical[18-23] studies have shown that the multiphoton Breit-Wheeler (BW) process[24], which was first experimentally demonstated at SLAC[25], is an effective route to generate high-energy dense positron beams. However, it is generally very difficulty to obtain all-optical dense GeV positron sources from direct laser-plasma interactions at a low laser intensity of $\sim 10^{22} \text{W cm}^{-2}$, together with controllable beam properties and feature of quasi-neutral pair flows. To the best of our knowledge, one of the unique physical properties of positron beams, the beam angular momentum (BAM), has not yet been considered in laser-plasma interaction to date.

High-energy particle beams with angular momentum have an additional degree of freedom and unique characteristics, which offer exciting and promising new tools for potential applications[2, 26-28] in test of the fundamental physical mechanics, probe of the particle property, generation of vortex beam, etc. For long time ago, it was realized that the circularly-polarized (CP) light beam might behavior like an optical torque[29], the spin angular momentum (SAM) of photons of such light can be transferred to the BAM of particles[30, 31]. As the light intensity increases, this effect could be significant via laser-driven electron acceleration[26, 32] and subsequent photon emission[27, 33, 34]. However, to the best of our knowledge, it has not yet been reported so far on how to obtain high-energy dense positron and γ-photon beams with a highly controllable angular momentum at currently affordable laser intensities.

In this paper, by use of ultra-intense CP lasers interacting with a near-critical-density (NCD) plasma, we show that ultra-bright GeV γ-photon and dense positron beams with controllable angular momentum can be efficiently produced via nonlinear Compton scattering (NCS)[35, 36] and the multiphoton BW process. It involves the effective transfer of the SAM of laser photons to the BAM of high-energy γ-photons and dense positrons, and offers a promising approach to manipulate such ultra-relativistic particle beams with special structure. This γ-photon and positron beams would provide a new degree of freedom to enhance the physics capability for applications, such as discovering new particles and unraveling the underlying physics via the collision of $e^+e^-/\gamma\gamma$. This scheme may provide possibilities for interdisciplinary studies[1, 2, 13], such as revealing the situations of rotating energetic systems, exploring fundamental QED processes, modelling astrophysical phenomena, and so on.

**RESULTS AND DISCUSSION**

**Overview of the scheme**

Figure 1a illustrates schematically our scheme and some key features of produced γ-photons and positron beams obtained based upon full three-dimensional particle-in-cell (3D-PIC) simulations with QED and collective plasma effects incorporated. In the first stage, electrons are accelerated by the drive laser and emit abundant γ-photons via the NCS. This results in strong radiation reaction (RR) forces[37-42], which act on the electrons so that a large number of electrons are trapped in the laser fields under the combined effect of the laser ponderomotive force and self-generated electromagnetic field. Dense helical beams of the trapped electrons and γ-photons are formed synchronously, as shown in Fig. 1b. As the trapped electrons collide with the opposite-propagating scattering laser in the second stage, the γ-photon emission is boosted significantly in number and energy. Finally, such bright γ-photons collide with the scattering laser fields to trigger the nonlinear BW process in the third stage. A plenty of GeV positrons with high density and controllable angular momentum are produced (Figs. 1c and 1d).

When the trapped electrons collide head-on with the opposite-propagating scattering laser pulse, the electrons undergo a strong longitudinal ponderomotive force and are reversely pushed away by such scattering laser, inducing a strong longitudinal positive current $J_x > 0$. Finally, the created pairs can be effectively separated from the background plasmas, so that quasi-neutral pair plasma is generated, as shown in Fig. 2. Since the ratio $D_{pair}/l_s \approx 1.8$ here, it is likely to induce collective effects of the pair plasma[9, 10], where $D_{pair}$ and $l_s = c/\sqrt{8\pi e^2 n_{e+}/\bar{\gamma} m_{e+}}$ are respectively the transverse size and the skin depth of the pair plasma, $\bar{\gamma}$ is the average Lorentz factor of the pair, $e$ is the elementary charge, $n_{e+}$ and $m_{e+}$ are the positron density and mass. Such dense GeV pair flows would offer new possibilities for future experimental study of the pair plasma physics in a straightforward and efficient all-optical way. This scheme could be thus used as a test bed for pair plasma physics and nonlinear QED effects, and may serve as compact efficient GeV positron and γ-ray sources for diverse applications.

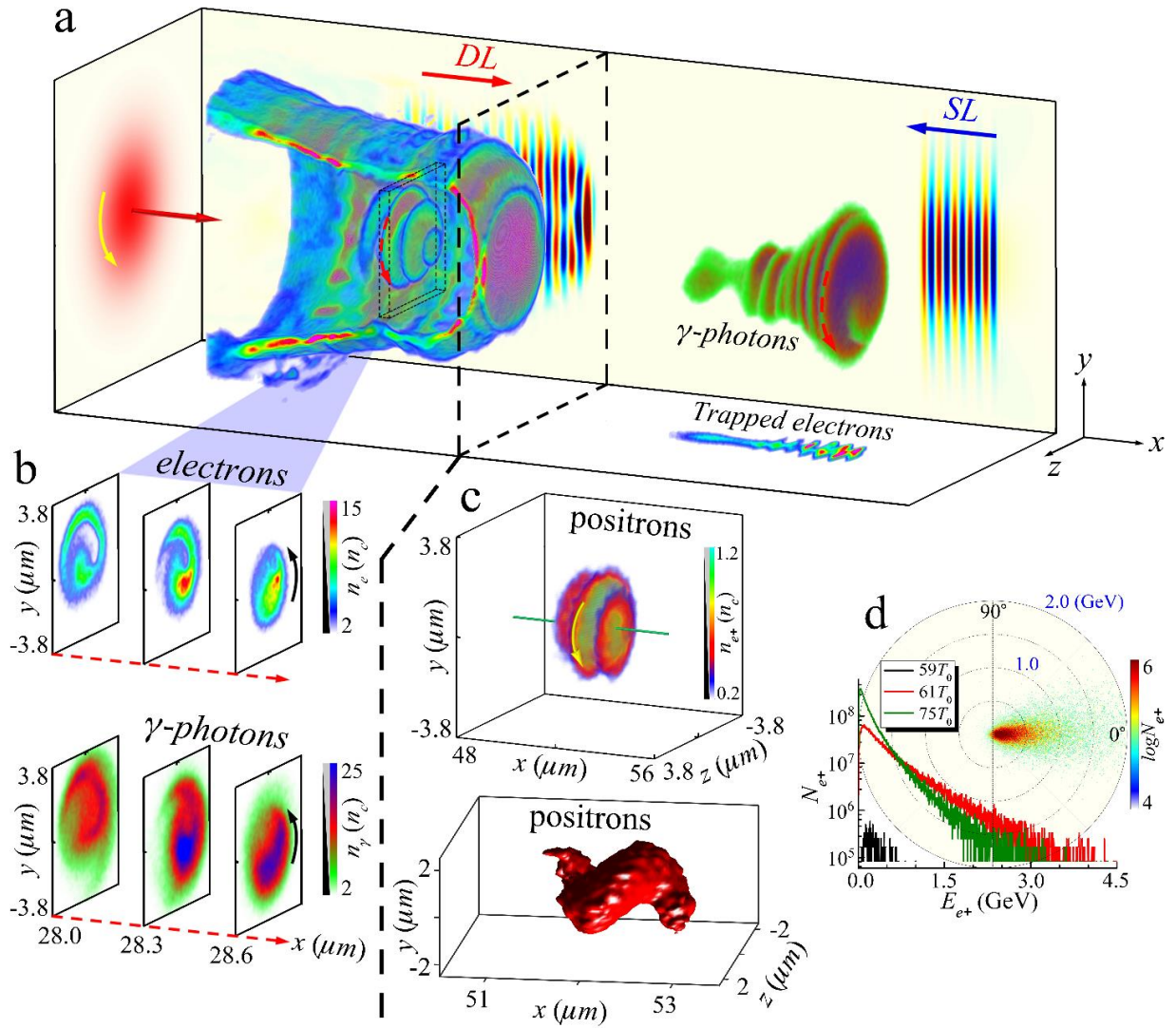

**Figure 1.** Helical particle beam generation via laser-plasma interactions. (**a**) Sketch of generating high-energy dense particle beams with high angular momentum in laser-NCD plasma interactions. The scheme is divided into three steps: (1) A right-handed CP drive laser (DL) pulse incident from the left irradiates a NCD plasma slab to generate and trap

a helical electron beam; (2) By colliding of the electron beam with a scattering laser (SL) pulse, dense energetic γ-photons are emitted via NCS; (3) The γ-photons further collide with the SL fields, which triggers the multiphoton BW process to produce numerous electron-positron pairs. The red- and blue-arrows indicate respectively the propagating directions of the DL and SL pulses. The black-dashed lines mark the dividing line between the step (1) and steps (2, 3). (**b**) Transverse slices of the density distributions of the trapped electrons and γ-photons at $t = 38T_0$. (**c**) The positron density distribution in 3D (top) and 3D isosurface distribution of positron energy density of $250n_c$MeV (bottom) at $t = 62T_0$. (**d**) The angular distribution of positron energy and snapshots of the positron energy spectrum at different times. The curved arrows indicate the rotation directions, while the green line in (**c**) shows the rotation axis. The density of electrons, γ-photons and positrons is normalized by the critical density $n_c$. The simulation results here are obtained when the DL pulse is with right-handed CP and the SL pulse is with left-handed CP.

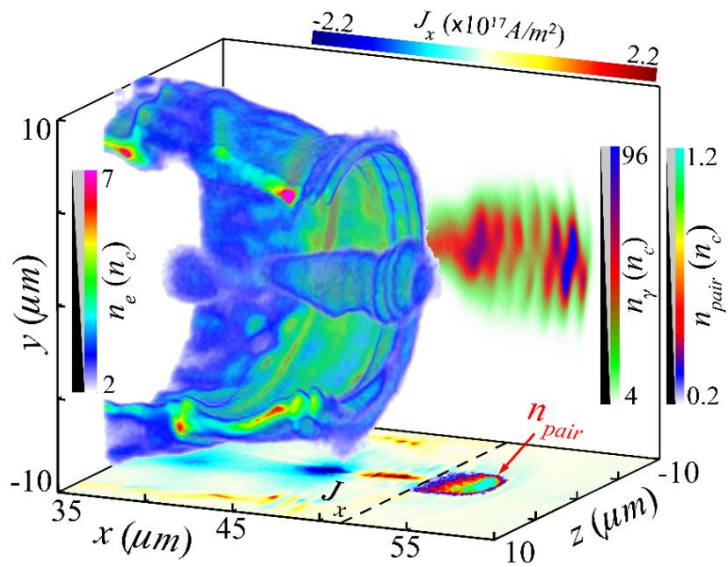

**Figure 2.** Achieving quasi-neutral pair plasma. Density distributions of electrons ($n_e$), γ-photons ($n_\gamma$ in the *x-y* plane), and pairs ($n_{pair}$ in the *x-z* plane) at $t = 65T_0$. The cross section of the longitudinal current ($J_x$) is projected in the *x-z* plane, where the black-dashed line marks the front of the trapped electron bunch. One can see clearly that the electron-positron pairs are separated from the background electrons and travel along the *x*-axis, resulting in quasi-neutral pair plasma formation.

**Numerical modelling**

Full 3D PIC **s**imulations have been carried out using the QED-PIC code EPOCH[43], where the effects of spin polarization[44, 45] are ignored. In the simulations, the drive laser pulse and the scattering laser pulse (with a delay of $45T_0$) are respectively incident from the left and right boundaries of the simulation box, possessing the same transversely Gaussian profile of $\exp(-r^2/r_0^2)$ ($r_0 = 4\lambda_0 = 4\mu m$ is the laser spot radius) and different longitudinally plateau profiles of $15T_0$ for the drive laser pulse and $5T_0$ for the scattering laser pulse. Their electric fields rotate in the same azimuthal direction, i.e., right-handed for the drive laser pulse and left-handed for the scattering pulse, respectively. The normalized laser amplitude is $a_0 = 120$, corresponding to currently approachable laser intensity[46] $\sim 10^{22}$ W cm$^{-2}$. The simulation box size is $x \times y \times z = 65\lambda_0 \times 20\lambda_0 \times 20\lambda_0$ with a cell size of $\Delta x = \lambda_0/25$ and $\Delta y = \Delta z = \lambda_0/15$, sampled by 16 macro-particles per cell. Absorbing boundary conditions are used for both fields and particles. To enhance the laser absorption in plasmas, a NCD hydrogen plasma slab located between $4\lambda_0$ to $50\lambda_0$ with an initial electron density of $n_0 = 1.5 n_c$ ($n_c = m_e \omega_0^2 / 4\pi e^2$, where $m_e$ is the electron mass, and $\omega_0$ is the laser angular frequency) is employed, which can be obtained from foam, gas or cluster jets[47, 48]. Such plasma density is transparent to the incident laser pulse due to the relativistic transparency effect $n_{c,\gamma} \sim a_0 n_c \gg n_0$, so that the drive laser pulse can propagate and accelerate electrons over a longer distance in the plasma. The parameters of lasers and plasmas are tunable in simulations.

**The QED emission model**

The stochastic emission model is employed and implemented in the EPOCH code using a probabilistic Monte Carlo algorithm[49, 50]. The QED emission rates are determined by the Lorentz-invariant parameter[51]: $\eta = (e\hbar/m_e^3 c^4)|F_{\mu\nu} p^\nu|$, which dominates the quantum radiation emission, and $\chi = (e\hbar^2/2m_e^3 c^4)|F_{\mu\nu} k^\nu|$, which dominates the pair production. Here $F_{\mu\nu}$ is the electromagnetic field tensor and the absolute value of the four-vector indicated by $|...|$, $p^\nu (\hbar k^\nu)$ is the electron's (photon's) four-momentum. In the laser-based fields, $\eta$ and $\chi$ can be expressed as $\eta = \frac{\gamma_e}{E_s}\sqrt{(\mathbf{E} + \boldsymbol{\beta} \times \mathbf{B})^2 - (\boldsymbol{\beta} \cdot \mathbf{E})^2}$, and $\chi = \frac{\hbar\omega}{2m_e c^2 E_s}\sqrt{(\mathbf{E} + \hat{\mathbf{k}} \times \mathbf{B})^2 - (\hat{\mathbf{k}} \cdot \mathbf{E})^2}$, where $\hbar\omega (\hbar k)$ and $\hat{\mathbf{k}}$ are the energy (momentum) and unit wave vector of the emitted photon, $\hbar$ is the reduced Planck constant, $E_s = m_e^2 c^3/e\hbar$ is the Schwinger electric field[52], $\boldsymbol{\beta} = \mathbf{v}/c$ is the normalized velocity of the electron by the speed of light in vacuum $c$, $\gamma_e$ is the electron relativistic factor, $\mathbf{E}$ and $\mathbf{B}$ are the electric and magnetic fields. The processes of photon and pair production have a differential optical depth[50] $d\tau_\gamma/dt = (\sqrt{3}\alpha_f c\eta)/(\lambda_C \gamma_e)\int_0^{\eta/2} F(\eta,\chi)/\chi d\chi$ and $d\tau_\pm/dt = (2\pi\alpha_f c/\lambda_C)(m_e c^2/\hbar\omega)\chi T_\pm(\chi)$, respectively, where $\alpha_f$ is the fine-structure constant, $\lambda_C$ is the Compton wavelength, $F(\eta,\chi)$ is the photon emissivity and $T_\pm(\chi)$ is the pair emissivity. The emission rate is solved until its optical depth is reached and simultaneously the emission occurs. When the energetic electrons and emitted γ-photons propagate parallel to the laser pulse, the contribution from the electric field on the parameters can be almost entirely canceled out by the magnetic field, resulting in $\eta \to 0$ and $\chi \to 0$. On the contrary, if the particles counter-propagate with the pulse, one can obtain $\eta \sim 2\gamma_e|\mathbf{E}_\perp|/E_s$ and $\chi \sim (\hbar\omega/m_e c^2)|\mathbf{E}_\perp|/E_s$, where $\mathbf{E}_\perp$ is the electric field perpendicular to the motion direction of the particles. The energy of most emitted photons can be predicted as[15] $\hbar\omega \approx 0.44\eta\gamma_e m_e c^2$. The parameter $\chi$ is thus rewritten as $\chi \sim 0.22\eta^2$ at the collision stage, implying that $\chi \propto \gamma_e^2 |\mathbf{E}_\perp|^2/E_s^2$ increases significantly with the laser intensity and the positron production becomes more efficient accordingly.

The radiating electrons experience a strong discontinuous radiation recoil (called the quantum-corrected RR force) due to the stochastic photon emission in the QED regime[49, 50]. The stochastic nature of such radiation allows electrons to attain higher energies and emit higher energy photons than those undergoing a continuous recoil in the classical regime. The resulting photon energies are comparable to the electron energies, as illustrated in Figs. 3a and 3d.

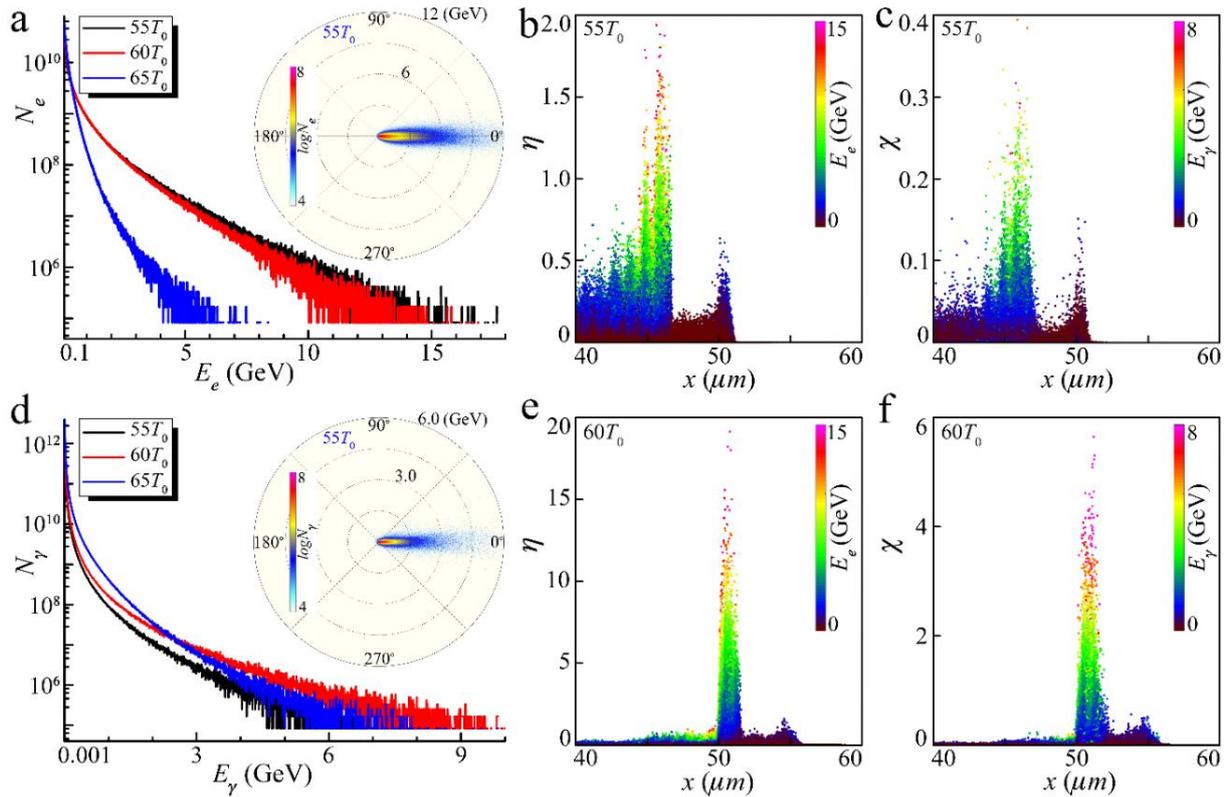

**Figure 3.** Particle energy spectra and two key quantum parameters of the QED effects. The energy spectra of electrons (**a**) and γ-photons (**d**) at $t = 55T_0$, $t = 60T_0$ and $t = 65T_0$. The corresponding insets in (**a**) and (**d**) show the angular-energy distribution at $t = 55T_0$. The spatial distributions of $\eta$ (**b**, **e**) and $\chi$ (**c**, **f**) along the x-axis: at $t = 55T_0$ (**b**, **c**) before the collision and $t = 60T_0$ (**e**, **f**) after the collision.

**Relativistic electron dynamics in laser-driven NCD plasma**

We can use the single electron model to understand the electron acceleration in NCD plasma. For a CP laser propagating along the x-direction, the components of laser fields are respectively $E_{Ly} = E_L \sin\phi$, $E_{Lz} = E_L \cos\phi$, $B_{Ly} = -E_{Lz}/v_{ph}$, $B_{Lz} = E_{Ly}/v_{ph}$, where $E_L$ is the amplitude of laser electric field, $\phi = kx - \omega_0 t$ and $v_{ph} = \omega_0/k$ are the laser phase and phase velocity, and $k$ is the wave number. The definition of self-generated azimuthal magnetic field is $\mathbf{B}_{S\theta} = B_{Sy}\hat{\mathbf{e}}_y + B_{Sz}\hat{\mathbf{e}}_z$, where $\hat{\mathbf{e}}_y$ and $\hat{\mathbf{e}}_z$ are the unit vector in the y and z direction, respectively. With considering the RR effect in laser-plasma interactions, the equation of motion of electrons can be expressed as $\frac{d\mathbf{p}}{dt} = \mathbf{f}_L + \mathbf{f}_{rad}$, where $\mathbf{p} = \gamma_e m_e \mathbf{v}$ is the electron momentum. The Lorentz force is $\mathbf{f}_L = -e(\mathbf{E} + \mathbf{v} \times \mathbf{B})$, where $\mathbf{E} \approx \mathbf{E}_L$ and $\mathbf{B} \approx \mathbf{B}_L + \mathbf{B}_{S\theta}$ are the local electric and magnetic fields. The RR force is considered as $\mathbf{f}_{rad} \approx -2eE_s\alpha_f\eta^2 g(\eta)\boldsymbol{\beta}/3$ by taking into account of quantum effects for ultra-relativistic limit $\gamma_e \gg 1$, where $g(\eta) = \left(3\sqrt{3}/2\pi\eta^2\right)\int_0^\infty F(\eta,\chi)d\chi$ is a function accounting for the correction in the radiation power excited by QED effects[51]. The transverse motion of the electron can be thus described as

$$\frac{dp_y}{dt} = -e\kappa E_L \sin\phi + ev_x B_{Sz} - F v_y, \tag{1}$$

$$\frac{dp_z}{dt} = -e\kappa E_L \cos\phi - ev_x B_{Sy} - F v_z, \tag{2}$$

where $\kappa = 1 - v_x/v_{ph}$ and $F = 2eE_s\alpha_f\eta^2 g(\eta)/3c$. For a highly relativistic electron, one can reasonably assume that $\dot{v}_x \approx 0$, $\dot{\gamma}_e \approx 0$, and $\dot{F} \approx 0$, because they are slowly-varying compared with the fast-varying term of $\dot{p}_y$ and $\dot{p}_z$. Under these assumptions, one can obtain the time derivative of the equations of electron motion as follow

$$\frac{d^2 p_y}{dt^2} + \left(\frac{ev_x}{\gamma_e m_e}\frac{\partial}{\partial r}B_{S\theta}\right)p_y + \left(\frac{F}{\gamma_e m_e}\right)\frac{dp_y}{dt} = a_0 m_e c \omega_L^2 \cos\omega_L t, \tag{3}$$

$$\frac{d^2 p_z}{dt^2} + \left(\frac{ev_x}{\gamma_e m_e}\frac{\partial}{\partial r}B_{S\theta}\right)p_z + \left(\frac{F}{\gamma_e m_e}\right)\frac{dp_z}{dt} = a_0 m_e c \omega_L^2 \sin\omega_L t. \tag{4}$$

Here $\frac{\partial}{\partial r}B_{S\theta} = \frac{\partial}{\partial z}B_{Sy} = -\frac{\partial}{\partial y}B_{Sz}$, $a_0 = eE_L/m_e c \omega_0$, and $\omega_L = \kappa\omega_0$ is the laser frequency in the moving frame of the electron. The resulting self-generated annular magnetic field is approximately proportional to the distance $r$ from the laser axis, so that $\frac{\partial}{\partial r}B_{S\theta}$ can be assumed as a constant in this case. The solutions of electron transverse momentum with time can be thus expressed as $p_{y,z}(t) = p_\perp \cos(\omega_L t + \phi_0)$, where $\phi_0$ is the initial phase and $p_\perp$ is the transverse momentum amplitude. By substituting $p_{y,z}(t)$ into Eqs. (3) and (4), one can obtain $p_\perp \propto \frac{a_0 m_e c \omega_L^4}{\omega_L^2 \omega_F^2 + (\omega_L^2 - \omega_{SB}^2)^2}$. Here $\omega_{SB} = \sqrt{\frac{ev_x}{\gamma_e m_e}\frac{\partial}{\partial r}B_{S\theta}}$ indicates the oscillation frequency induced by the self-generated annular magnetic field, and $\omega_F = \frac{F}{\gamma_e m_e}$ indicates the oscillation frequency induced by the RR effect. In the ultra-relativistic limit $\gamma_e \gg a_0 \gg 1$, one has $\omega_{SB}/\omega_L \propto 1/\sqrt{\gamma_e} \ll 1$ and $\omega_F/\omega_L \propto a_0 f_{rad}/eE_L \gamma_e \ll 1$, so that the transverse momentum amplitude is approximately given by $p_\perp \propto a_0 m_e c$. The transverse momentum of the electron can be thus expressed as $p_y \propto a_0 m_e c \cos\omega_L t$ and $p_z \propto a_0 m_e c \sin\omega_L t$. In such highly collimated relativistic case ($p_x \approx \gamma_e m_e c$ and $p_x^2 \gg p_\perp^2$), one can obtain $\gamma_e = (a_0^2 + \Gamma^2 + 1)/2\Gamma \propto \sim a_0^2/2\Gamma$, where $\Gamma$ is a small constant. Thus the energy of accelerated electrons is proportional to the square of laser amplitude, as shown in Fig. 4a.

The instantaneous radiation power induced by the electron can be written as $f_{rad}v \approx 2eE_s c\alpha_f \eta^2 g(\eta)/3$. For $\eta \ll 1$, the RR effect can be ignored ($f_{rad} \ll f_L$), but it becomes significant and the quantum radiation needs to be considered for $\eta \sim O(1)$. In our configuration, $\eta$ can easily reach unity (see Figs. 3b and 3e), leading to $f_{rad} \sim O(f_L)$. Thus the RR effect dominates and QED effects come into play, so that most electrons are efficiently trapped and accelerated inside the laser pulse instead of being expelled away, constituting a dense electron beam. Figure 4b shows the self-generated annular magnetic field $B_{S\theta} \sim 2\pi r e n_e$. This plays a significant role in electron trapping by the centripetal pinching force $e\boldsymbol{\beta} \times \mathbf{B}_{S\theta} \sim O(eE)$ and in the γ-photon emission by the quantum radiation $\gamma_e |\boldsymbol{\beta} \times \mathbf{B}_{S\theta}|/E_s \to \eta \sim O(1)$[53, 54]. Thus, the electron trapping is attributed to the RR effect together with the self-generated magnetic field around the channel dug by the CP laser in the NCD plasma. Here the NCD plasma is employed for providing more background electrons being trapped and accelerated, and for forming an intense self-generated magnetic field, which is very beneficial for the high-energy γ-photon emission.

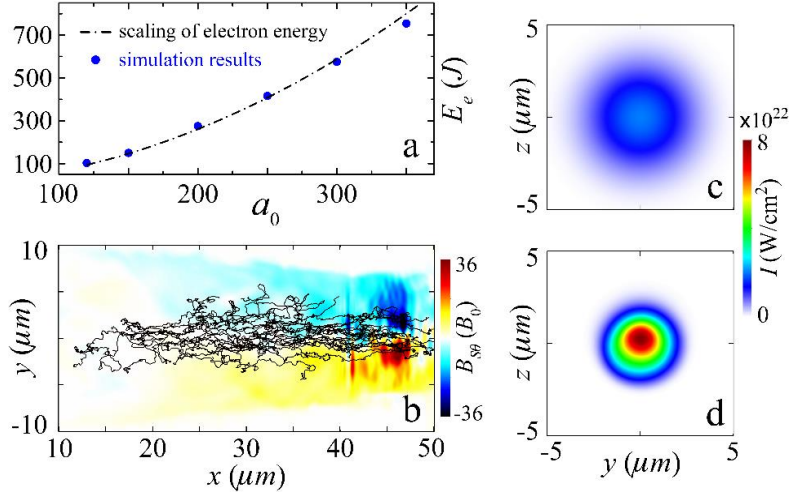

**Figure 4.** (**a**) The energy evolution of trapped electrons as the laser amplitude. The blue dots show the simulation results, and the black dotted curve presents the $a_0^2$ scaling of electron energy. (**b**) The black lines plot the trajectories of some electrons from the trapped energetic electron beam. The background color map shows the self-generated azimuthal magnetic field $B_{S\theta}$, normalized by $B_0 = m_e \omega_0/e$. Spatial distributions of the laser intensity in the y-z plane for the incident laser pulse (**c**) in vacuum and (**d**) in plasma channel.

When an ultra-intense Gaussian laser pulse propagates in the NCD plasma[53, 55], electrons can be accelerated directly by the laser fields, which induces intense electric currents and forms plasma lens. This enhances the laser intensity greatly (see Figs. 4c and 4d) due to the strong relativistic self-focusing and self-compression of the pulse in plasmas, the resulting laser amplitude is significantly boosted with the dimensionless parameter $a_{foc} \sim 250$. In the stochastic photon emission, electrons can be accelerated directly by the laser to a high energy without radiation loss or radiation recoil before they radiate high-energy photons. The maximum energy of electrons is thus approximately proportional to the focused laser amplitude $a_{foc}^2$, which energy can be as high as 10 GeV. For most energetic electrons, their energies are significantly decreased due to the high-energy γ-photon emission when the RR effect comes into play. Finally, a substantial energy of the radiating electrons is converted into high-energy γ-photons, providing an efficient and bright GeV γ-ray source.

For effective γ-photon emission, here we only consider the trapped electrons with energy >100MeV from the region within the off-axis radius $< 4\mu m$, whose total number can be estimated as $N_e \sim n_e \pi R_e^2 d_e$. Here $R_e \sim (\lambda_0/\pi)\sqrt{a_0 n_c/n_e}$ is the electron beam radius, $d_e \sim ct_0$ is the beam length (matching with the laser pulse duration), $t_0$ is the formation time of the electron beam dependent on the laser-plasma interaction. Thus the total electron number is $N_e \sim a_0 n_c \lambda_0^2 ct_0/\pi$.

**Efficient generation and control of high angular momentum of electron, γ-photon and positron beams**

In our scenario, the NCS at the collision stage of energetic electrons with the scattering laser pulse dominates the radiation emission over that at the first stage, resulting in greatly boosted γ-photon emission, as seen in Fig. 5a. In the regime of quantum-dominated radiation production, the total radiation power of the trapped electrons can be estimated by

$$P_{rad} = \Sigma_i (f_{rad,i} v_i) \approx \frac{2}{3} N_e e E_s c \alpha_f \overline{\eta^2 g(\eta)}, \tag{5}$$

where the subscript $i$ indicates serial number of individual electrons, $N_e$ is the electron number, $\overline{\eta^2 g(\eta)}$ is the average radiation power factor of trapped energetic electrons with the parameter $\eta_i > 0.1$ determining the photon emission. The radiation power predicted by Eq. (5) at the first and second stages is $P_{rad} \approx$ 0.6PW and 6.8PW, respectively, in good agreement with the PIC simulation results of 0.4PW and 7PW. This suggests that the NCS in NCD plasma provides an efficient way to generate GeV γ-rays in the multi-PW level with extremely high intensity of $\sim 10^{23}$W cm$^{-2}$.

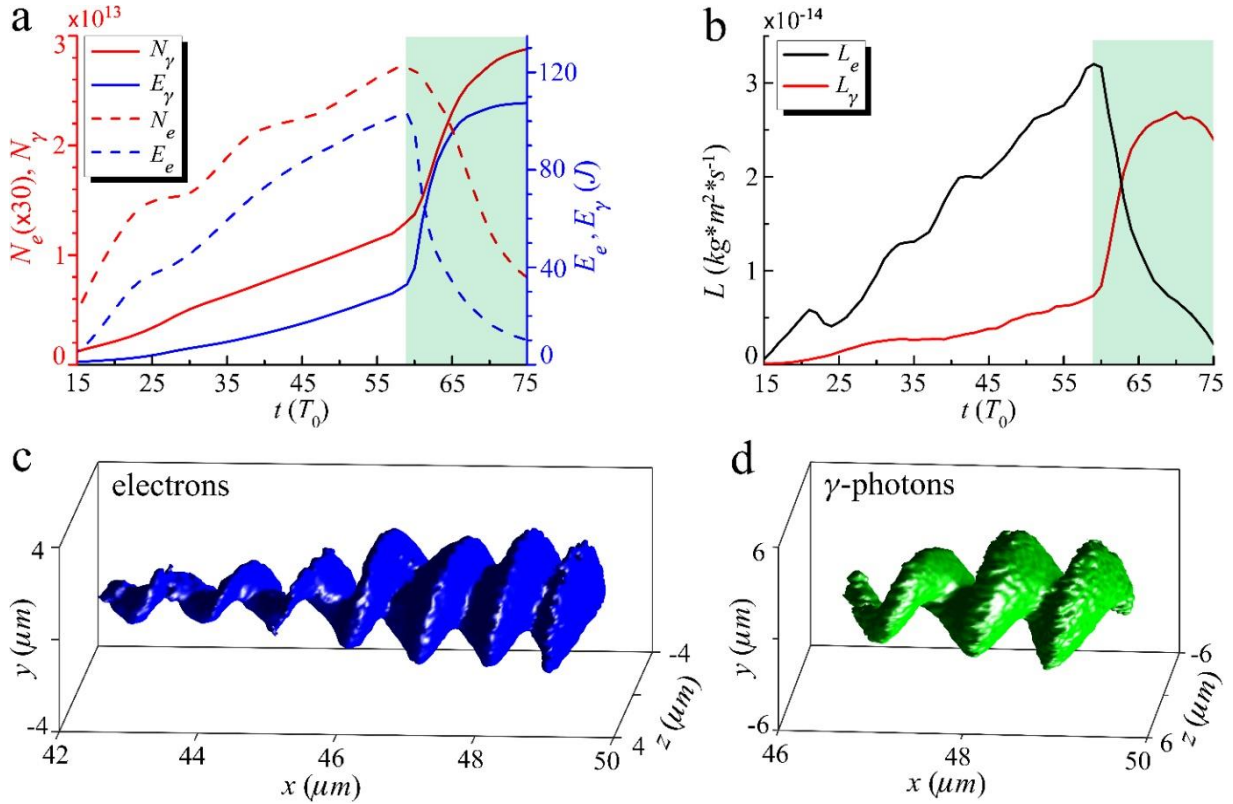

**Figure 5.** (**a**) Evolution of the γ-photon yield (red-solid line) and energy (blue-solid line), and the trapped electron number (red-dashed line) and energy (blue-dashed line). (**b**) The angular momentum evolution of the trapped electrons (black line) and γ-photons (red line). The green shadows in (**a**) and (**b**) indicate entering the collision stage. Plots in (**c**) and (**d**) illustrate 3D isosurface distributions of the energy density of electrons and γ-photons at $t = 58T_0$,

respectively, where the isosurface values are $2000n_c$MeV and $500n_c$MeV, respectively. The simulation parameters are the same as in Fig. 1.

In addition to the extreme high peak power of γ-rays, the twist formation of the γ-photon beam is particularly interesting, which is found to be strongly dependent upon the polarization of the laser pulse or the SAM of laser photons. For a drive laser pulse propagating along the $x$ direction, generally its electric field can be described as $\mathbf{E}_L = E_L[\sqrt{2-\delta^2}\sin(\phi)\hat{\mathbf{e}}_y + \delta\cos(\phi)\hat{\mathbf{e}}_z]$, $\delta$ determines the polarization state: $\delta = 0$ for linearly-polarized (LP) laser; $\delta = \pm 1$ for right-handed and left-handed CP laser, respectively; $0 < |\delta| < 1$ for elliptically-polarized (EP) laser. One can obtain $p_y \propto \sqrt{2-\delta^2}\cos(\phi)a_0 m_e c$ and $p_z \propto \delta\sin(\phi)a_0 m_e c$. The electron transverse position with respect to the laser axis is given by $r_\perp = \int p_\perp/(\gamma m_e)dt$. Thus, the total angular momentum of electrons in such laser fields can be approximately expressed as

$$L_e = \sum_i |\mathbf{r}_{\perp i} \times \mathbf{p}_{\perp i}| \propto \delta\sqrt{2-\delta^2}a_0 m_e c^2 N_e/\omega_0, \tag{6}$$

where $N_e \sim a_0 n_c \lambda_0^2 c t_0/\pi$ is the number of the trapped electrons before entering the collision stage. This implies that the electron angular momentum can be optically controlled by tuning the laser parameters, e.g., $L_e \sim 0$ when $\delta = 0$ for LP lasers and it becomes maximum when $\delta = 1$ for CP lasers. Since $N_e$ scales almost linearly with the interaction time $t_0$, the electron angular momentum increases accordingly, which is substantiated by the PIC simulations as exhibited in Fig. 5b. For such an isolated system, both the momentum and angular momentum of particles and photons conserve during the laser-plasma interaction, so that electrons can obtain their angular momentum from the CP laser pulses with SAM and then transfer it to the γ-photons via the NCS. Figure 5b also presents the evolution of the angular momentum of emitted γ-photons, which shows a clear transfer of the angular momentum from the electrons to the γ-photons at the second stage. The 3D isosurface distributions of the angular momenta for the electron beam and the γ-photons in Figs. 5c and 5d illustrate the clear helical structures, which have right-handedness following the laser and have the period of about one laser wavelength. This suggests that the γ-photons can bear angular momenta up to $\sim 2.7 \times 10^{-14} \text{kg} \cdot m^2 \cdot s^{-1}$. Furthermore, the γ-photons have an unprecedented peak brightness of $\sim 10^{24}$ photons/s/mm$^2$/mrad$^2$/0.1%BW at 100 MeV with a laser energy conversion efficiency of 8.1%. With the forthcoming lasers such as ELI[11] with intensity of $10^{23}$Wcm$^{-2}$, the BAM of γ-photons is as high as $8 \times 10^{-13} \text{kg} \cdot m^2 \cdot s^{-1}$, which corresponds to $10^7 \hbar$ per photon in average with a total photon yield of $\sim 6 \times 10^{14}$.

Although the γ-photons emitted at the first stage have a large number and high energies, these γ-photons almost co-move with the drive laser pulse so that the key quantum parameter $\chi$ determining the positron generation becomes much smaller, i.e., $\chi \lesssim O(0.1)$. This results in very limited positron production. At the subsequent stage when the γ-photons collide with the scattering laser pulse, $\chi$ is greatly increased ($\chi \sim 4$, as seen in Fig. 3f), leading to highly efficient positron production via the multiphoton BW process. Finally, a high-yield ($2.5 \times 10^{10}$) dense ($\sim n_c$) GeV positron beam with the same helicity of right-handedness as the γ-photon beam is produced, as shown in Fig. 1c.

In our configuration, the drive laser pulse acts as micro-motors to seed angular momentum in laser-NCD plasma interactions, while the scattering laser pulse triggers the multiphoton BW process and plays the role of a torque regulator. The positron angular momentum is mainly controlled by tuning the polarization of the drive laser pulse. To illustrate this, we consider two drive laser pulses with different polarization: $\delta = 0$ for LP; and $\delta = \sqrt{2}/2$ for EP, while the scattering laser is always left-handed CP. According to Eq. (6), the electron angular momentum is determined by $\delta$, implying that it can reach the maximum $L_e^{CP}$ when $\delta = 1$, while $L_e^{LP} \to 0$ for $\delta = 0$ and $L_e^{EP} \to \sqrt{3} L_e^{CP}/2$ for $\delta = \sqrt{2}/2$. The simulation results show a good agreement with the theoretical predictions: $L_e^{LP} \approx 0$ by a LP laser, and $L_e^{EP} \approx 0.82 L_e^{CP}$ by the given EP laser. Meanwhile, the angular momentum of the γ-photons changes similarly, because their angular momentum originates from the parent electrons, as plotted in Fig. 6b. We also see that the positrons can get less angular momenta from the scattering laser pulse ($\lesssim 0.17 L_{e^+}^{CP}$, as plotted in Fig. 6c) when the LP drive laser is used. This is reasonable because the parent electrons and γ-photons obtain almost no angular momentum from the LP drive laser pulse. This offers an efficient and straightforward approach to explore the angular momentum transfer between the laser and particles, and to optically control the angular momentum and helicity of such energetic beams for future studies.

One can also tune the angular momentum of positrons by simply changing the chirality of the scattering laser pulse. This can be attributed to the torsional effect of the CP laser (like a wrench). The electric fields of both drive laser and scattering laser pulses rotate in the same azimuthal direction when using a left-handed CP scattering laser pulse, whereas their electric fields rotate in the opposite direction when using a right-handed CP scattering laser pulse, so that they can tune the rotation of the positron beam in the same direction or in the opposite direction, respectively. As a result, the positron angular momentum tuned by the left-handed CP scattering laser pulse is higher than that tuned by the right-handed CP scattering laser pulse, as seen in Fig. 6a.

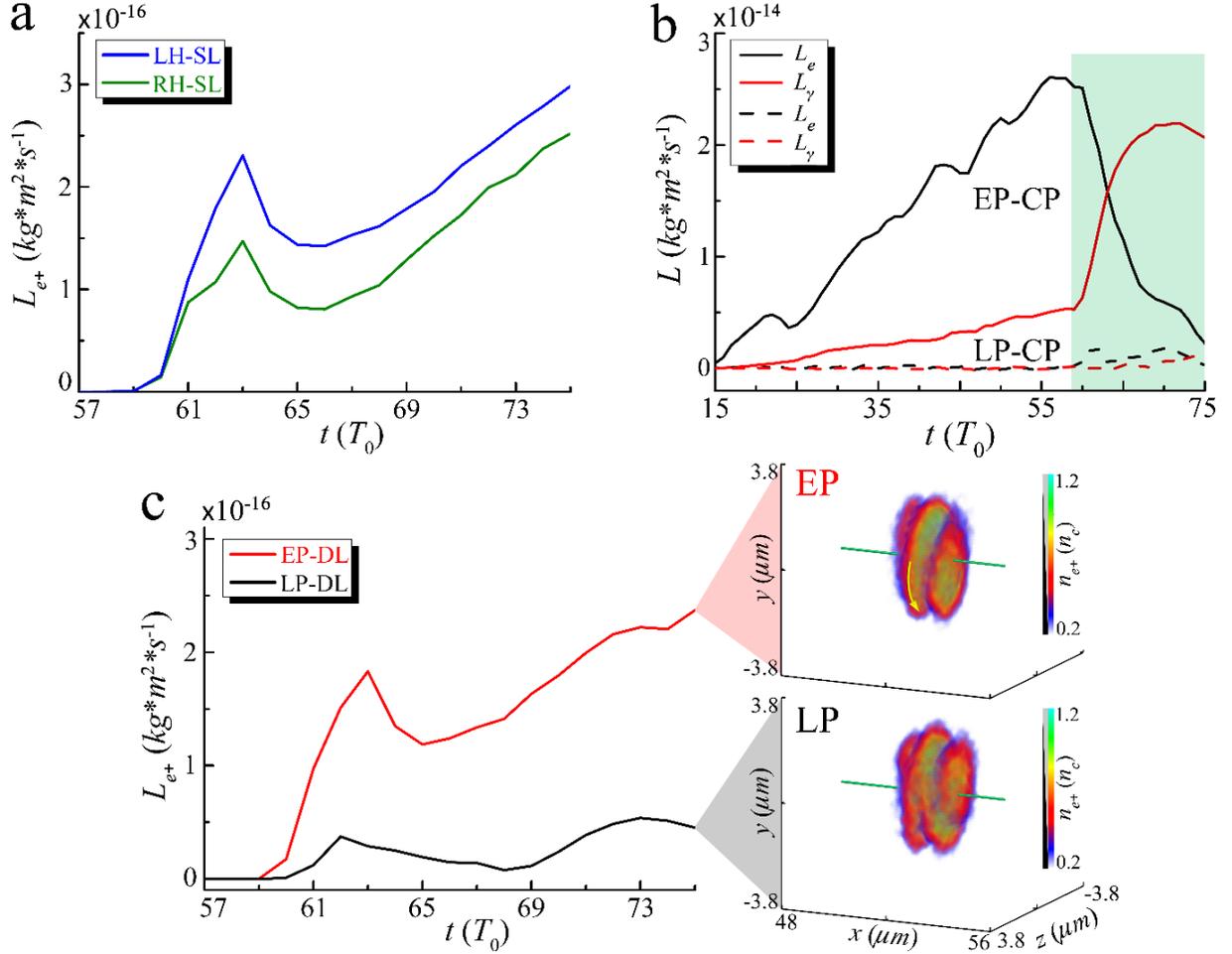

**Figure 6.** Optical control of the angular momentum of the γ-photons and positrons. (**a**) Temporal evolution of the angular momentum of positrons using a right-handed CP drive laser (DL) pulse and CP scattering laser (SL) pulses incident from the right side with different helicities: RH, right-handed (green line); LH, left-handed (blue line). Temporal evolution of the angular momentum of trapped electrons and γ-photons (**b**) and positrons (**c**) using a left-handed CP SL pulse and DL pulses with different polarizations. In (**b**), dashed line for LP laser with $\delta = 0$ and solid line for right-handed EP laser with $\delta = \sqrt{2}/2$, where the green shadow indicates entering the collision stage. In (**c**), the angular momentum is shown for DL pulses with $\delta = \sqrt{2}/2$ (EP), and 0 (LP). The insets in (**c**) show the positron density distributions in the cases of using EP (top) and LP (bottom) DL pulses, respectively, at $t = 62T_0$.

**Robustness of the regime and discussion**

Further simulations have been performed to explore the scaling of the angular momentum of energetic particle beams as a function of the laser amplitude, as shown in Fig. 7a. Here, we keep the parameter $a_0 n_c/n_0 = \text{constant}$ for ultra-relativistic laser-plasma interactions[56]. The efficiency of positron production increases with the laser intensity, which is valid for all considered laser intensities. In our configuration, the key quantum parameter $\chi \sim O(1)$, so that the multiphoton BW process can be efficiently triggered to create copious numbers of high-energy electron-positron pairs. However, $\chi$ is greatly suppressed at a much lower laser intensity, e.g, $\chi \ll 1$ at $a_0 < 100$, which leads to very limited positron creation. This indicates that there exsits a threshold laser amplitude (i.e., $a_{th} \sim 120$ in our scenario), for highly-efficient prolific positrons production.

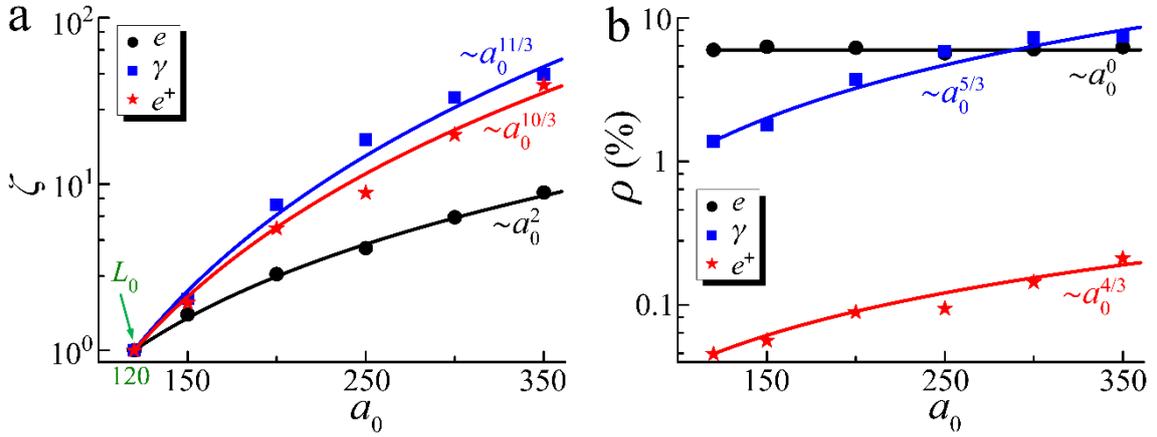

**Figure 7.** Scaling of the angular momentum and conversion efficiency of energetic particle beams. (**a**) The angular momentum scaling of electrons $\zeta_e$ (black line), γ-photons $\zeta_\gamma$ (blue line), and positrons $\zeta_{e^+}$ (red line), defined as the ratio $\zeta = L/L_0$, as a function of the dimensionless laser amplitude $a_0$, where $L_0$ is the angular momentum of electrons, γ-photons, or positrons measured at $a_{th} = 120$. (**b**) The angular momentum conversion efficiency from the laser pulse with SAM to electrons ($\rho_{l \to e}$, black line), γ-photons ($\rho_{l \to \gamma}$, blue line), and positrons ($\rho_{l \to e^+}$, red line), as a function of $a_0$.

For lasers with a fixed polarization, we all obtain $L_e \propto a_0^2$ from Eq. (6). The ratio scaling of the electron angular momentum is then expressed as $\zeta_e = L_e/L_{e,0} \sim (a_0/a_{th})^2$ for different laser intensities. This is in accordance with the simulation results, as seen in Fig. 7a. The instantaneous radiation power emitted by the electron scales as $\eta^2 g(\eta)$ and the photon angular momentum can be then approximately written as $L_\gamma \propto L_e \eta^2 g(\eta) \propto a_0^{\sim 10/3}$. The positron angular momentum originates mainly from the γ-photons so that its scaling shows a similar trend. Finally, we obtain that $\zeta_\gamma \sim (a_0/a_{th})^{11/3}$ and $\zeta_{e^+} \sim (a_0/a_{th})^{10/3}$ from the PIC simulations, which is reasonably close to the analytical estimation above.

Here, the total SAM of the CP drive laser pulse approximates $L_l = \delta\hbar N_l$, where $N_l = E_l/\hbar\omega_0 \propto a_0^2$ is the total laser photon number. Thus the conversion efficiency from the laser SAM to the angular momentum of electrons, γ-photons, and positrons scales: $\rho_{l\rightarrow e} \sim a_0^0$, $\rho_{l\rightarrow\gamma} \sim a_0^{5/3}$, and $\rho_{l\rightarrow e^+} \sim a_0^{4/3}$, respectively. The scaling is well validated by the simulation results as shown in Fig. 7b. Taking the forthcoming lasers like ELI for example, we estimate the positron angular momentum up to $6.5 \times 10^{-15} \text{kg} \cdot m^2 \cdot s^{-1}$ with a high angular momentum conversion efficiency of ~0.15%. Since the γ-photons are emitted by electrons in helical motion and via the NCS of a CP pulse, the resulting high angular momentum γ-rays may carry away well-defined orbital angular momentum along the propagation direction, which has been verified theoretically in recent several studies[27, 28]. By manipulating the electron motion in plasmas, it is a potential way for generating γ-ray beams with coherent angular momentum. Such γ-ray beams may find potential applications, such as providing additional effects in the interaction with nucleus, controlling nuclear processes, probing the structure of particles, etc.

**CONCLUSION**

In conclusion, we have proposed and numerically demonstrated an efficient scheme to produce high angular momentum γ-photon and positron beams in NCD plasma irradiated by two counter-propagating high power lasers with circular polarization under currently affordable laser intensity $\sim 10^{22}$W cm$^{-2}$. It is shown that ultra-intense multi-PW γ-rays and dense ($10^{21}$cm$^{-3}$) GeV positrons with a high charge number of ~4 nano-Coulombs can be efficiently achieved. In addition, the angular momentum and helicity of such energetic beams are well controlled by the drive laser pulse and are tunable by the scattering laser pulse. With the upcoming laser facilities like ELI, this all-optical scheme not only provides a promising and practical avenue to generate ultra-bright PW γ-rays and dense positron beams with GeV energies and high angular mometum for various applications, but also enables future experimental tests of nonlinear QED theory in a new domain.

**ACKNOWLEDGEMENTS**

This work was partially supported by the National Basic Research Program of China (Grant Nos. 2013CBA01504 and 2018YFA0404802), the National Natural Science Foundation of China (Grant Nos. 11721091, 11774227, 11622547, and 11655002), the Science and Technology Commission of Shanghai

Municipality (Grant No. 16DZ2260200), the Ministry of Science and Technology of China for an International Collaboration Project (Grant No. 2014DFG02330), Hunan Provincial Natural Science Foundation of China (Grant No. 2017JJ1003), Fok Ying-Tong Education Foundation (Grant No. 161007), and a Leverhulme Trust Grant at the University of Strathclyde. The EPOCH code was in part developed by the UK EPSRC grant EP/G056803/1. All simulations have been carried out on the PI supercomputer at Shanghai Jiao Tong University.

Laser with aWakefield-Accelerated Electron Beam. *Phys. Rev. X* **2**, 041004 (2012).

40. Ji, L. L., Pukhov, A., Kostyukov, I. Y., Shen, B. F. & Akli, K. Radiation-reaction trapping of electrons in extreme laser fields. *Phys. Rev. Lett.* **112**, 145003 (2014).

41. Zhang, P., Ridgers, C. P. & Thomas, A. G. R. The effect of nonlinear quantum electrodynamics on relativistic transparency and laser absorption in ultra-relativistic plasmas. *New J. Phys.* **17**, 043051 (2015).

42. Gonoskov, A. *et al.* Ultrabright GeV Photon Source via Controlled Electromagnetic Cascades in Laser-Dipole Waves. *Phys. Rev. X* **7**, 041003 (2017).

43. Arber, T. D. *et al*. Contemporary particle-in-cell approach to laser-plasma modelling. *Plasma Phys. Control. Fusion* **57**, 113001 (2015).

44. King, B., Elkina, N. & Ruhl, H. Photon polarization in electron-seeded pair-creation cascades. *Phys. Rev. A.* **87**, 042117 (2013).

45. Del Sorbo, D. *et al.* Spin polarization of electrons by ultraintense lasers. *Phys. Rev. A.* **96**, 043407 (2017).

46. Bahk, S.-W. *et al*. Generation and characterization of the highest laser intensities ($10^{22}$ W/cm$^2$). *Opt. Lett.* **29**, 2837 (2004).

47. Ma, W. *et al*. Directly synthesized strong, highly conducting, transparent single-walled carbon nanotube films. *Nano Lett.* **7**, 2307 (2007).

48. Fukuda, Y. *et al*. Energy increase in multi-MeV ion acceleration in the interaction of a short pulse laser with a cluster-gas target. *Phys. Rev. Lett.* **103**, 165002 (2009).

49. Duclous, R., Kirk, J. G. & Bell, A. R. Monte Carlo calculations of pair production in high-intensity laser-plasma interactions. *Plasma Phys. Control. Fusion* **53**, 015009 (2011).

50. Ridgers, C. P. *et al*. Modelling gamma-ray photon emission and pair production in high-intensity laser–matter interactions. *J. Comput. Phys.* **260**, 273 (2014).

51. Ritus, V. I. Quantum effects of the interaction of elementary particles with an intense electromagnetic field. *J. Russ. Laser Res.* **6**, 497–617 (1985).

52. Schwinger, J. On Gauge Invariance and Vacuum Polarization. *Phys. Rev.* **82**, 664 (1951).

53. Zhu, X. L. *et al*. Enhanced electron trapping and γ ray emission by ultra-intense laser irradiating a near-critical-density plasma filled gold cone. *New J. Phys.* **17**, 053039 (2015).
**20 / 21**